\documentclass[prl,aps,twocolumn,floats,epsf,superscriptaddress]{revtex4}  

\usepackage{epsfig}
\usepackage{dcolumn}  
\usepackage{relsize}

\begin{document}
\preprint{\tighten\vbox{\hbox{\hfil BELLE-CONF-0417}
                        \hbox{\hfil ICHEP04 11-0663}
}}

\newcommand{\bb}{\ensuremath{b\overline{b}}}
\newcommand{\BB}{\ensuremath{B\overline{B}}}
\newcommand{\Bbar}{\ensuremath{\overline{B}}}
\newcommand{\Bbaro}{\ensuremath{\overline{B}{}^0}}
\newcommand{\Bbarp}{\ensuremath{B{}^+}}
\newcommand{\Bbarm}{\ensuremath{B{}^-}}
\newcommand{\de}{\ensuremath{\Delta E}}
\newcommand{\mb}{\ensuremath{M_{\mbox{\scriptsize bc}}}}
\newcommand{\etap}{\ensuremath{\eta^\prime}}
\newcommand{\fbm}{\, \mbox{fb\ensuremath{^{-1}}}} 
\newcommand{\BF}{\ensuremath{{\mathcal B}}}
\newcommand{\dbar}{\ensuremath{\overline{d}}}
\newcommand{\LK}{\ensuremath{{\cal L}}}
\newcommand{\RK}{\ensuremath{{\cal R}_K}}
\newcommand{\FD}{\ensuremath{{\cal F}}}
\newcommand{\LR}{\ensuremath{{\cal R_L}}}
\newcommand{\hel}{\ensuremath{{\cal H}}}
\newcommand{\cost}{\ensuremath{\cos\theta_T}}
\newcommand{\cosb}{\ensuremath{\cos\theta_B}}
\newcommand{\sperp}{\ensuremath{S_\perp}}
\newcommand{\epp}{\ensuremath{\eta' \to \eta \pi^+ \pi^-}}
\newcommand{\erg}{\ensuremath{\eta' \to \rho^0 \gamma}}
\newcommand{\ebeam}{\ensuremath{E_{\mbox{\scriptsize beam}}}}
\newcommand{\babar}{BABAR}
\newcommand{\cs}{\ensuremath{/c^2}}

\vspace*{1ex}

\title{\large
       Observation of $\Bbaro\to D^0 \etap$ and $\Bbaro\to D^{*0} \etap$}

\date{\today}

\affiliation{Budker Institute of Nuclear Physics, Novosibirsk}
\affiliation{Chiba University, Chiba}
\affiliation{Chonnam National University, Kwangju}
\affiliation{University of Cincinnati, Cincinnati, Ohio 45221}
\affiliation{University of Frankfurt, Frankfurt}
\affiliation{Gyeongsang National University, Chinju}
\affiliation{University of Hawaii, Honolulu, Hawaii 96822}
\affiliation{High Energy Accelerator Research Organization (KEK), Tsukuba}
\affiliation{Hiroshima Institute of Technology, Hiroshima}
\affiliation{Institute of High Energy Physics, Chinese Academy of Sciences, Beijing}
\affiliation{Institute of High Energy Physics, Vienna}
\affiliation{Institute for Theoretical and Experimental Physics, Moscow}
\affiliation{J. Stefan Institute, Ljubljana}
\affiliation{Kanagawa University, Yokohama}
\affiliation{Korea University, Seoul}
\affiliation{Kyungpook National University, Taegu}
\affiliation{Swiss Federal Institute of Technology of Lausanne, EPFL, Lausanne}
\affiliation{University of Ljubljana, Ljubljana}
\affiliation{University of Maribor, Maribor}
\affiliation{University of Melbourne, Victoria}
\affiliation{Nagoya University, Nagoya}
\affiliation{Nara Women's University, Nara}
\affiliation{National Central University, Chung-li}
\affiliation{National United University, Miao Li}
\affiliation{Department of Physics, National Taiwan University, Taipei}
\affiliation{H. Niewodniczanski Institute of Nuclear Physics, Krakow}
\affiliation{Nihon Dental College, Niigata}
\affiliation{Niigata University, Niigata}
\affiliation{Osaka City University, Osaka}
\affiliation{Osaka University, Osaka}
\affiliation{Panjab University, Chandigarh}
\affiliation{Peking University, Beijing}
\affiliation{Princeton University, Princeton, New Jersey 08545}
\affiliation{Saga University, Saga}
\affiliation{University of Science and Technology of China, Hefei}
\affiliation{Seoul National University, Seoul}
\affiliation{Sungkyunkwan University, Suwon}
\affiliation{University of Sydney, Sydney NSW}
\affiliation{Tata Institute of Fundamental Research, Bombay}
\affiliation{Toho University, Funabashi}
\affiliation{Tohoku Gakuin University, Tagajo}
\affiliation{Tohoku University, Sendai}
\affiliation{Department of Physics, University of Tokyo, Tokyo}
\affiliation{Tokyo Institute of Technology, Tokyo}
\affiliation{Tokyo Metropolitan University, Tokyo}
\affiliation{Tokyo University of Agriculture and Technology, Tokyo}
\affiliation{University of Tsukuba, Tsukuba}
\affiliation{Virginia Polytechnic Institute and State University, Blacksburg, Virginia 24061}
\affiliation{Yonsei University, Seoul}
 \author{J.~Sch\"umann}\affiliation{Department of Physics, National Taiwan University, Taipei} 
  \author{K.~Abe}\affiliation{High Energy Accelerator Research Organization (KEK), Tsukuba} 
  \author{K.~Abe}\affiliation{Tohoku Gakuin University, Tagajo} 
  \author{H.~Aihara}\affiliation{Department of Physics, University of Tokyo, Tokyo} 
  \author{Y.~Asano}\affiliation{University of Tsukuba, Tsukuba} 
  \author{T.~Aushev}\affiliation{Institute for Theoretical and Experimental Physics, Moscow} 
  \author{S.~Bahinipati}\affiliation{University of Cincinnati, Cincinnati, Ohio 45221} 
  \author{A.~M.~Bakich}\affiliation{University of Sydney, Sydney NSW} 
  \author{Y.~Ban}\affiliation{Peking University, Beijing} 
  \author{I.~Bedny}\affiliation{Budker Institute of Nuclear Physics, Novosibirsk} 
% ...
\collaboration{The Belle Collaboration}

\tighten

\begin{abstract}
We report the observation of $\Bbaro \to D^0 \etap$ and the first  
observation of $\Bbaro \to D^{*0} \etap$, 
using 140\fbm{} of data collected at the $\Upsilon (4S)$ resonance with 
the Belle detector at the KEKB asymmetric energy $e^+ e^-$ 
collider. We find the branching fractions to be 
$\BF(\Bbaro\to D^0 \etap ) = [ 1.14 \pm 0.20$ (stat) 
$^{+0.10}_{-0.13}$ (syst)]$ \times 10^{-4}$ and 
$\BF(\Bbaro \to D^{*0} \etap ) = [ 1.21 \pm 0.34$ (stat) 
$\pm 0.22$ (syst)]$ \times 10^{-4}$ 
with significances including systematic uncertainty of $8.9$ and $5.3$ 
standard deviations, respectively.
\end{abstract}

\maketitle
 

The colour-suppressed decays $\Bbaro \to D^{(*)0} \, h^0$ have been observed 
for
$h^0=\pi^0,\eta,\omega$ and $\rho^0$~\cite{Belle1,Belle2,CLEO,BaBarD} 
and recently also $\Bbaro \to D^0 \, \etap$~\cite{BaBarD}.
%
The rates are all larger than originally expected in generalized
factorization~\cite{Neubert}. The large branching fractions may be
explained by final state rescattering~\cite{Hou01} or non-factorizable
diagrams in perturbative QCD~\cite{Lue03}. Hard-collinear effective
theory~\cite{Mantry04} cannot explain the rate for $\Bbaro \to
D^{(*)0} \, \pi^0$; other decay modes have not been investigated in
this framework.

\begin{figure}[!htb]
\centerline{
\epsfxsize 3.0 truein \epsfbox{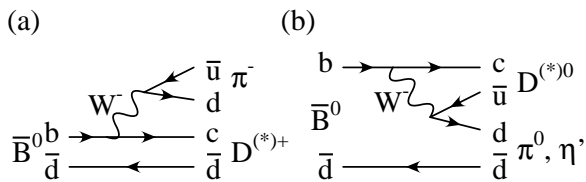}
}
\caption{\label{fig:feyn} Colour-allowed (a) and colour-suppressed (b)
tree level Feynman diagrams for $\Bbar \to D^{(*)} h$.}
\end{figure}

Both the colour-allowed decay $\Bbaro \to D^{(*)+} h^-$~\cite{CC} and the
colour-suppressed decay $\Bbaro \to D^{(*)0} h^0$ proceed via the
emission of a virtual $W^-$. In the former case the $W^-$ can decay without
colour restrictions,
however, in the latter case the $d$ quark from the $W^-$ 
decay has to carry
colour that matches the colour of the $\dbar$ quark originating from the
$\Bbaro$. This allows colour-singlet formation only in one out of three
cases resulting in suppression by a factor of $1\over{9}$ (therefore
colour-suppressed). The tree level diagrams for both decays are shown
in Fig.~\ref{fig:feyn}.
The contribution of the $W$-exchange diagram is usually
assumed to be negligible~\cite{Neubert}.

The branching fraction for $\Bbaro \to D^0 \etap$ and an upper limit 
for the 
$\Bbaro \to D^{*0} \etap$ decay have recently been published by
BaBar~\cite{BaBarD}.
In this Letter we present the first observation of
$\Bbaro \to D^{*0} \etap$ and an observation of $\Bbaro \to D^0 \etap$ with
more than $5 \Sigma$ statistical significance.

This analysis is based on a $140$\fbm{} data sample containing $152\times 10^6$
\BB{} events collected with the Belle detector at the KEKB~\cite{KEKB} $e^+ e^-$
collider. KEKB is an asymmetric energy collider 
($3.5$ GeV on $8$ GeV) that
operates at the $\Upsilon(4S)$ resonance ($\sqrt{s} = 10.58$ GeV) with peak
luminosity of nearly $1.4\times 10^{-34}$ cm$^{-2}$s$^{-1}$. $B^+B^-$ and
$B^0\Bbaro$ pairs are assumed to be produced with equal rates. For signal and
background simulation the QQ~\cite{QQ} Monte Carlo (MC) generator was used and 
the GEANT3~\cite{GEANT} package was used for detector simulation.

The Belle detector is a large-solid-angle magnetic
spectrometer consisting of a three-layer silicon vertex detector (SVD),
a 50-layer central drift chamber (CDC), an array of
aerogel threshold \v{C}erenkov counters (ACC), 
a barrel-like arrangement of time-of-flight
scintillation counters (TOF), and an electromagnetic calorimeter
comprised of CsI(Tl) crystals (ECL) located inside 
a superconducting solenoid coil that provides a 1.5~T
magnetic field.  An iron flux-return located outside of
the coil is instrumented to detect $K_L^0$ mesons and to identify
muons (KLM).  The detector
is described in detail elsewhere~\cite{Belle}.

Charged tracks with impact parameters less than $0.5$ cm radially and less than
$2$ cm in $z$ (the $z$ axis is anti-parallel to the positron beam direction)
with respect to the interaction point
and with momenta larger than $100$ MeV/$c$ are selected. 
Kaon and pion mass hypotheses are assigned to charged tracks on
the basis of a likelihood $\LK_{K/\pi}$ that is 
obtained by combining information from the CDC ($dE/dx$), ACC and TOF systems.
The likelihood ratio $\RK = \LK_{K}/(\LK_{\pi}+\LK_{K})$ ranges 
between $0$ (pion-like) and $1$ (kaon-like) and we require 
$\RK > 0.6$ for kaon and $\RK <0.4$ for
pion candidates originating from the heavy meson yielding an efficiency of about
$88\%
$ in both cases and $\RK <0.9$ for pion
candidates from the light meson side corresponding to an efficiency of $99\%
$. In addition, we reject tracks that are
consistent with an electron or muon hypothesis.

Photon candidates are selected with a minimum energy requirement of $50$ MeV.
Neutral pion candidates are reconstructed by combining two photons with 
invariant mass between $115$ MeV/$c^2$ and $152$ MeV/$c^2$, which 
corresponds to $\pm 2.5$
standard deviations in terms of the mass resolution. 
In addition, we require the momentum of the $\pi^0$ to be
greater than $400$ MeV/$c$ in the laboratory frame.

The $\eta$ meson is reconstructed by combining two photons. 
An asymmetric $\eta$ mass window was chosen, corresponding to $^{+2}_{-3}$
standard deviations. The mass of the
$\eta$ meson is then constrained to its nominal value from the Particle Data
Group (PDG)~\cite{PDG}.

We select $\rho$ candidates from two oppositely charged pion candidate
tracks in a mass region of $-220$ MeV\cs{} 
$< M(\pi^+ \pi^-) - m(\rho)< 170$ MeV\cs,
where $m_{\rho}$ is the nominal mass of $\rho^0$ mesons from the PDG and
$M(\pi^+ \pi^-)$ is the mass of the $\rho$ candidate. 
In addition, we 
require the transverse momentum of the daughter pions, $P_T(\pi)$, to be greater
than $300$ MeV/$c$. This requirement suppresses around $40 \% 
$ of the continuum background while retaining $86 \%
$ of the signal candidates. 

The \etap{} meson is reconstructed in its two dominant decay channels 
$\etap \to \eta \pi^+ \pi^-$ and $\etap \to \rho \gamma$. For both subdecays the
\etap{} candidates are required to be within a mass window of 
$0.94$ GeV\cs{} $< M(\etap) < 0.975$ GeV\cs{} and have a center-of-mass (CM)
momentum of
$P_{CM}(\etap) > 1.7$ GeV/$c$. The pions from the \etap{} 
decay must have a transverse momentum in the labratory frame of 
$P_T(\pi) > 0.1$ GeV/$c$. For the latter subdecay, the photon from the \etap{}
is required to have an energy greater than $0.2$ GeV in the lab frame. 
This selection alone suppresses around $85 \%
$ of background while retaining $85\%
$ of signal candidates.
A weak requirement on the helicity of the $\eta$ meson of $|h(\eta)| < 0.97$
is applied, where $h(\eta)$ is the cosine of the angle between the $\etap$
momentum and the direction of one of the decay photons
in the $\eta$ rest frame.
An \etap{} mass constraint is applied for the final fitting results.

The $D^0$ mesons are reconstructed in three different decay channels, $D^0 \to
K^-\pi^+$, $D^0\to K^-\pi^+\pi^0$ and $D^0\to K^-\pi^+\pi^-\pi^+$ 
with mode-dependent $\pm 3 \sigma$
mass windows. The $D^0 \to K^-\pi^+\pi^0$ decay channel has an
additional requirement on the $\pi^0$ momentum of $P(\pi^0)>0.4$ GeV$/c$ in 
the lab frame to suppress
continuum background. A $D^0$ mass constraint yields a significant
improvement to the momentum 
resolution for the $D^0 \to K^-\pi^+\pi^0$ decay and is applied for
all decay channels for consistency.

The $D^{*0}$ mesons are reconstructed combining a $D^0$ meson with a $\pi^0$ or
a $\gamma$.
We require the mass difference of the $D^0$ and $D^{*0}$ to satisfy $0.136$
GeV\cs{} $<M(D^{0}\pi^0)-M(D^0)<0.148$ GeV\cs{}
and $0.131$ GeV\cs{} $<M(D^{0}\gamma)-M(D^0)<0.156$
GeV\cs . In addition, we require the momentum of the $D^0$ in the lab 
frame 
to be greater than $0.8$ GeV$/c$.

The $\Bbaro$ mesons are reconstructed combining an $\etap$ meson and a $D^0$ or 
$D^{*0}$ meson. Two kinematic variables are used to extract the $\Bbaro$ meson
signal: the energy difference $\de = E_B-\ebeam$ and the beam constrained
mass $\mb =\sqrt{\ebeam^2 - P_B^2}$, where \ebeam{} is the beam energy and $E_B$
and $P_B$ are the reconstructed energy and momentum of the $\Bbaro$ 
candidate in the
$\Upsilon(4S)$ rest frame. The events that satisfy the requirements $\mb > 5.2$
GeV\cs{} and $|\de| <0.25$ GeV are selected for further analysis.

After these selection criteria, the two major background sources are continuum
$e^+e^- \to q \bar{q}$ (where $q = u,d,s,c$) and $b \to c$ decays.

Several event shape variables (defined in the CM frame) are used to distinguish
the more spherical \BB{} topology from 
the jet-like $q\bar q$ continuum events. 
The thrust angle $\theta_T$ is defined
as the angle between the primary $\Bbaro$ decay daughters
and the thrust axis formed by all tracks not from the same $\Bbaro$
meson.
Jet-like events tend to peak near $|\cost| = 1$,
while spherical events have a flat distribution.
The requirement $|\cost| < 0.9$ is applied prior to any other 
event topology selections. 

Additional continuum suppression is obtained by using modified 
Fox-Wolfram moments~\cite{SFW} and the angle $\theta_B$ between the flight 
direction of the reconstructed $\Bbaro$ candidate and the beam axis. A 
Fisher discriminant (\FD)~\cite{fisher} is formed by a linear combination
of $\cost$, $\sperp$ and five modified Fox-Wolfram moments. 
$\sperp$ is the ratio of
the scalar sum of the transverse momenta of all tracks outside a 
$45^{\circ}$ cone around the $\etap$ direction 
to the scalar sum of their total momenta.
Probability density functions (PDFs) are obtained from signal and 
background MC data samples.
These variables are then combined to form a topological likelihood function
$\LK_c = $PDF$_c(\cosb) \cdot $PDF$_c(\FD)$
where $c = $ signal ($s$) or continuum background $(q\bar{q})$.
Signal follows a $1-\cos^2(\theta_B)$ distribution while continuum background is
uniformly distributed in $\cosb$.
We select signal-like events by requiring a high likelihood ratio
$\LR = \LK_{s}/(\LK_{s} + \LK_{q\bar q})$ to 
suppress continuum background. For channels with an \erg{} decay an additional
variable $\cos\theta_{\hel}$, which is the angle between 
the $\etap$ momentum and the direction of one of
the decay pions in the $\rho$ rest frame, is
included for better signal-background separation. The \LR{} requirements 
are optimized
using signal and continuum background Monte Carlo samples and are found to be 
strongly mode dependent and ranging from $0.1$ to $0.925$, where in general more
restrictive
constraints are applied for decays with \erg{} and looser constraints for decays
including a $D^{*0}$. 

For events with multiple $\Bbaro$ candidates, the best candidate is
selected on the basis of the $\chi^2$ for the vertex fit to the charged pions
from the \etap . If multiple $D^{(*)}$ meson candidates remain after
this selection, the best candidate is selected using $\chi^2 = (M(D^{(*)}) -
m_{D^{(*)}} )^2 / \sigma^2$, where $M(D^{(*)})$ is the $D^{(*)}$ candidate mass,
$m_{D^{(*)}}$ is the nominal $D^{(*)}$ mass and $\sigma$ is the resolution of
reconstructed $D^{(*)}$ mass. 
Multiple candidates in the heavy meson side appear for less than $2\%
$ of the events for the $\Bbaro \to D^0 \etap$ decay but in the order of $20\%
$ for the $\Bbaro \to D^{*0} \etap$ decay.

Backgrounds from other $B$ decay modes such as $\Bbarm \to D^{(*)0} \rho^-$
and $\Bbarm \to D^{(*)0} a_1^-$ are studied. These decays are suppressed by
vetoing events where the $\rho$ or $a_1$ could be constructed from the $\Bbaro$
candidate daughters with $-0.1$ GeV$ < \de < 0.08$ GeV and $\mb >5.27$ GeV\cs.
The vetos suppress about $10\% -30\%
$ of the \BB{} background 
after the final selection criteria, while retaining over $98\%
$ of the signal MC events.
For decays with \epp , which have less background contamination,
the remaining \BB{} contributions are
modeled with a single two-dimensional smooth function, obtained from a large MC
sample. For decays with \erg{} we divide these backgrounds into
two groups, the $\Bbarm \to D^{(*)0} a_1^-$ decays and all others, both are
again modeled with two-dimensional smoothed histograms.

The signal PDFs are represented by a Gaussian plus a bifurcated Gaussian (a
Gaussian with different width on either side of the mean) in \de{}
and a bifurcated Gaussian in \mb. All PDFs are extracted from Monte Carlo
simulations. For $\Bbaro \to D^0 \etap$ events the feeddown from 
$\Bbaro \to D^{*0} \etap$ is modeled using $2$-dimensional smoothed histograms.
Continuum events are represented by a first or second order
polynomial in \de{} and an empirical background function introduced by
ARGUS~\cite{ARGUS} for \mb. 
Signals should peak around $\de = 0$ GeV and $\mb = 5.28$ GeV\cs. Correction 
factors
accounting for the difference between MC and data, determined from a study of
the high statistics decay mode $\Bbaro \to \etap K_S$, are applied to
the mean and width values of the signal shapes for both \de{} and \mb.

The reconstruction efficiencies are determined from a large sample of signal MC
events, and range from $0.6\% -11.2\%
$ for the 18 different decay channels, where the subdecay branching fractions
are not included. Correction factors due to differences
between data and MC are applied for the charged track
identification, photon, $\pi^0$ and $\eta$ reconstruction.

The signal yields ($N_S$) are extracted using extended unbinned 
maximum-likelihood fits simultaneously performed in \de{} and \mb. 
An extended likelihood function is:
\begin{eqnarray}
L(N_S,N_{B_j}) = \hspace*{5cm} \nonumber \\
\frac{e^{-(N_S+\sum_jN_{B_j})}}{N!} \prod_{i=1}^{N}
\left[N_{S} P_S(\vec x_i) + \sum_j N_{B_j} P_{B_j}(\vec x_i)\right]
\label{eq:ns-lkhd}
\end{eqnarray}
where $N$ is the total number of events, $i$ is and index running over the
events, $\vec x_i$ is a vector of the \de{} and \mb{} values for each event and
$P_S$ and $P_{B_j}$ are the probability density functions for
signal and background, respectively, and the index $j$ runs over 
all background sources. 
The signal yield $N_S$ and background contents $N_{B_j}$ are determined
by maximizing the $L(N_S,N_{B_j})$ function in the $(N_S,N_{B_j})$
manifold, where the $N_{B_j}$ defines a $j$-dimensional submanifold of all
different backgrounds. The statistical significance of the signal
is calculated as $\Sigma = \sqrt{2\ln(L_{\rm max}/L_0)}$,
where $L_{\rm max}$ and $L_0$ denote
the maximum likelihood value and the
likelihood value at zero branching fraction, respectively.

\begin{figure}[!htb]
\centerline{
\epsfxsize 1,7 truein \epsfbox{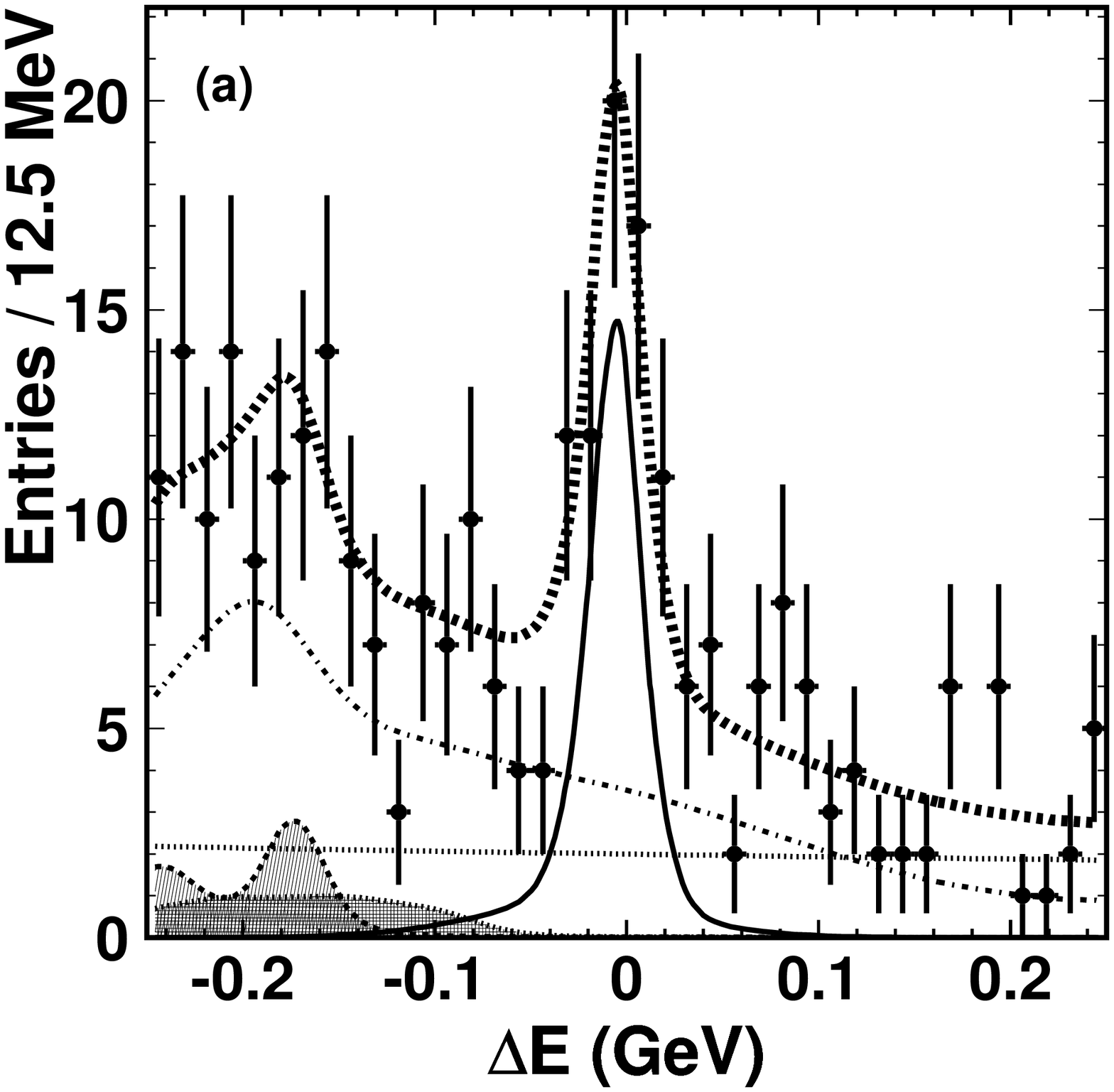}
\epsfxsize 1.7 truein \epsfbox{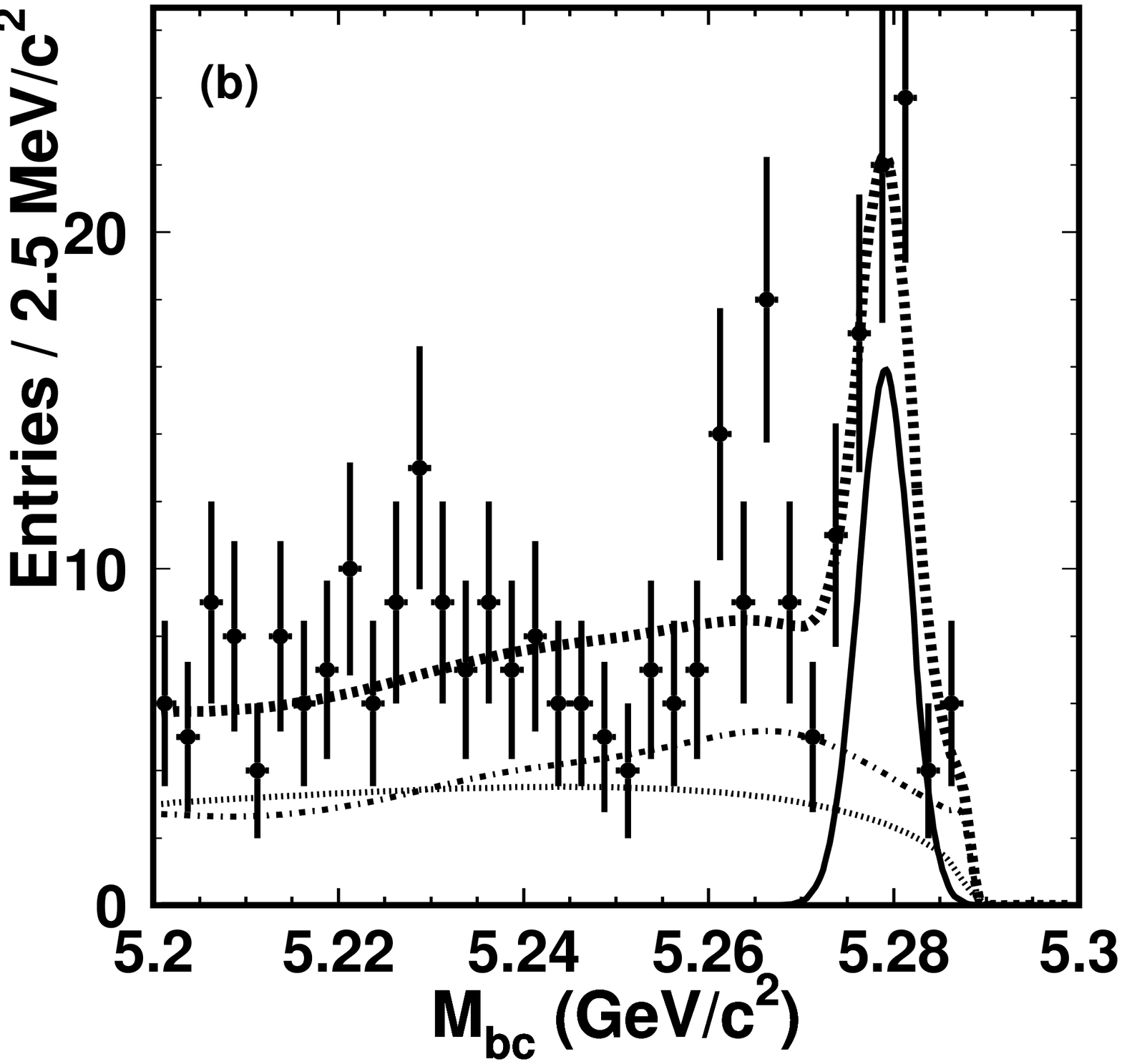}
}
\centerline{
\epsfxsize 1.7 truein \epsfbox{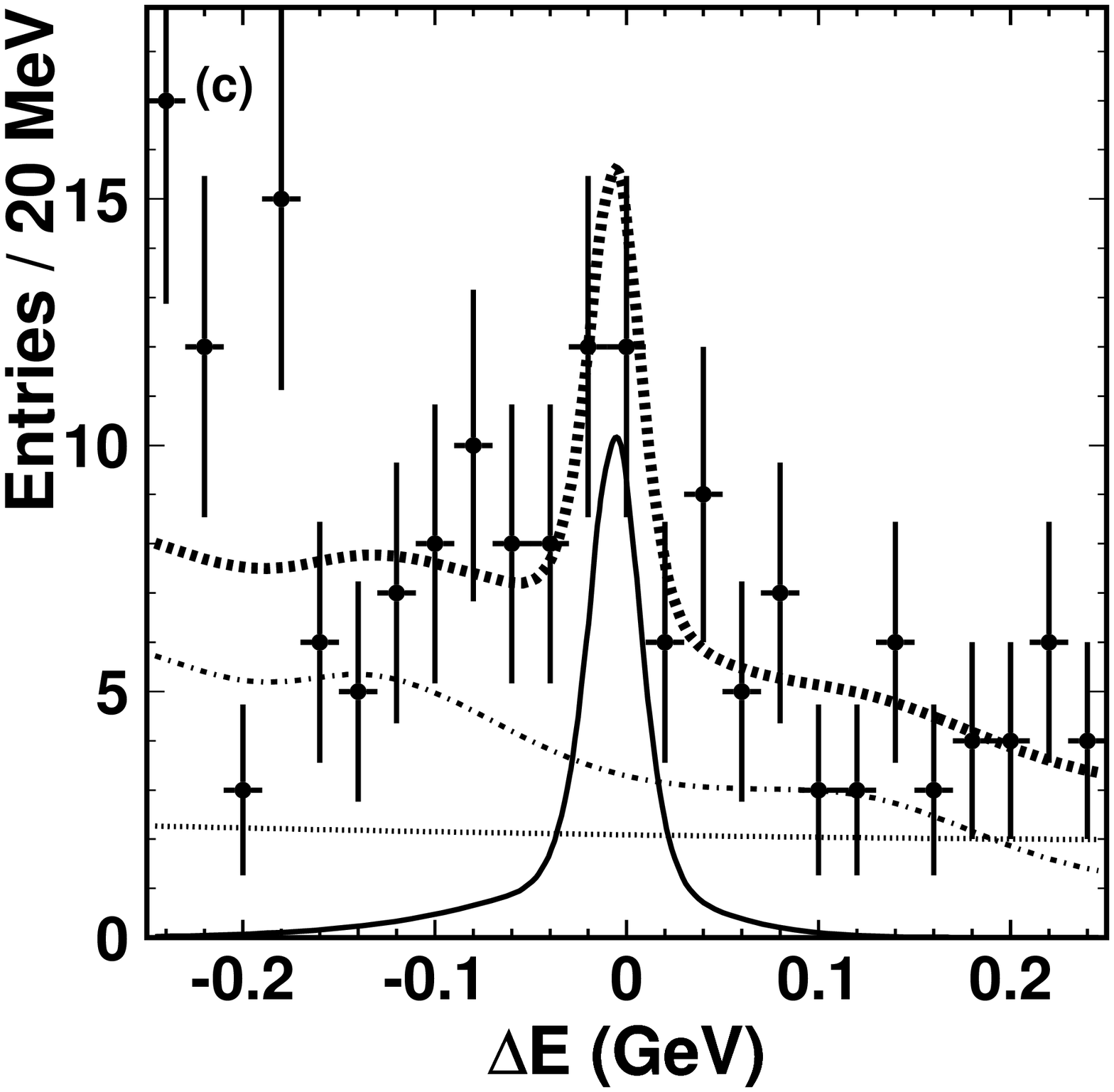}
\epsfxsize 1.7 truein \epsfbox{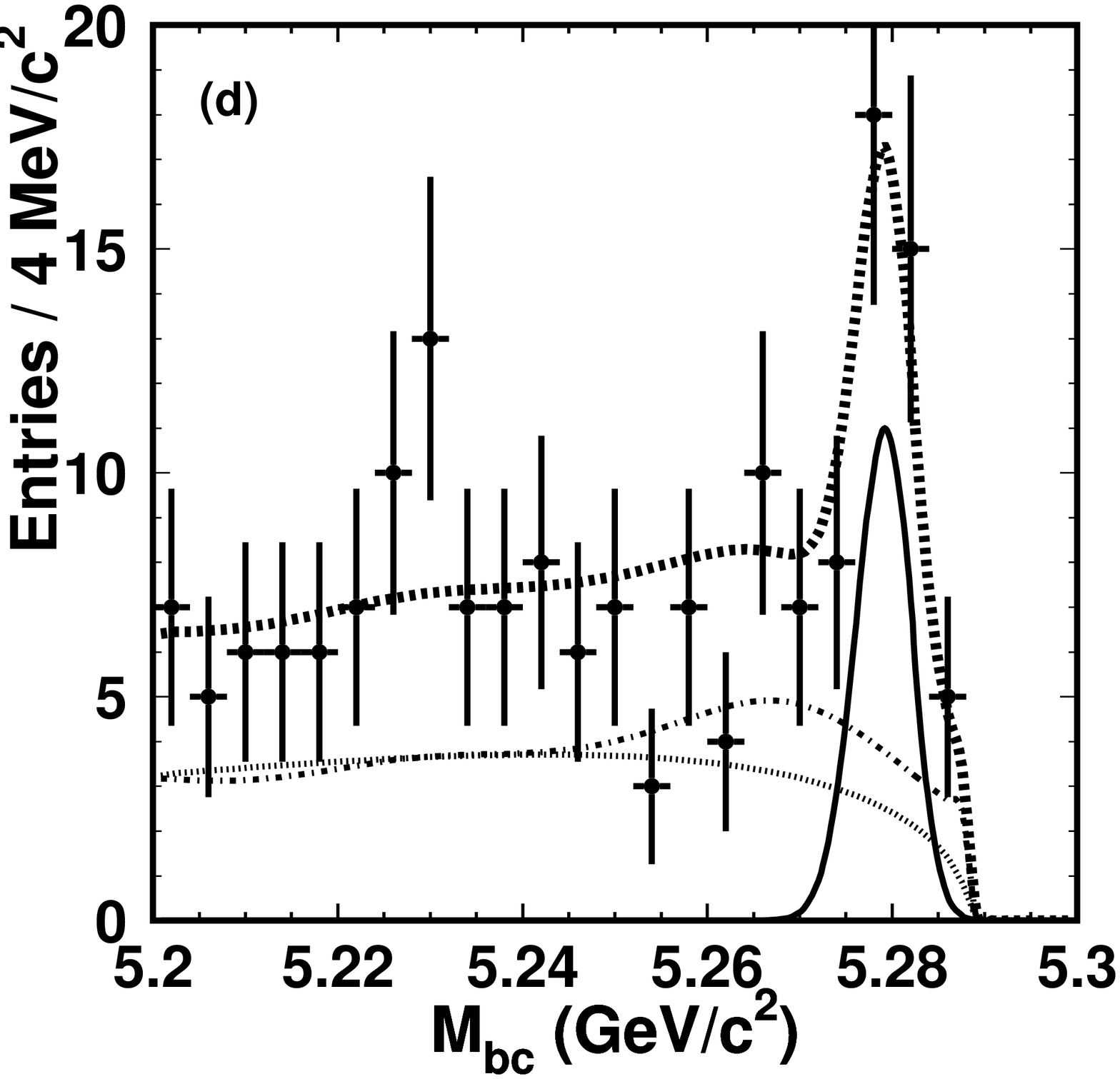}
}
\caption{\label{fig:resultsD} $\Bbaro \to D^0 \etap$ decay \de{} (a) and \mb{}
(b) projection plots and $\Bbaro \to D^{*0} \etap$ decay \de{} (c) and 
\mb{} (d) projection plots. 
Points with error bars represent the data, the solid line is the
signal contribution, the dash-dotted line is \BB{} and the dotted line
is continuum background. The two shaded areas represent the $D^{*0} \to D \pi^0$
(dashed line, filled area) and $D^{*0} \to D \gamma$ (dotted line, hatched area) 
feeddown. The dashed line is the sum of all contributions. The projection plots
are for signal regions of all other variables, the signal regions for \de{} and
\mb{} are: $-0.055$ GeV $<\de<0.05$ GeV and $5.27$ GeV $<\mb<5.29$ GeV.}
\end{figure}

The signal and background normalisations are floated in the fit while
other PDF parameters are fixed to values determined from MC studies.
When combining two or more modes we fit for a common branching fraction instead
of the signal yield for each mode.
In the case of $\Bbaro \to D^{0} \etap$, the contributions from
the $D^{*0}$ feeddown are fixed to the branching fraction obtained in
this analysis and the number of $D^{*0}$ events is calculated with
a mode dependent ``feeddown-efficiency'', which is derived from MC
simulations of $\Bbaro \to D^{*0} \etap$ decays reconstructed as
$\Bbaro \to D^{0} \etap$. The branching fractions together with the
yields, the reconstruction efficiency for each decay and the significances are
listed in Table~\ref{tab:results}. The \de{} and \mb{} projections of
the fits are shown in Fig.~\ref{fig:resultsD}.
\begin{table}[htbp]
\begin{center}
\caption{\label{tab:results}
Signal yields,
efficiencies $( = \sum_i \epsilon_i {\cal B}(D^{(*)}_i )$, with the index $i$
running over the $D^{(*)}$ subdecay modes), branching fractions ${\cal B}$ 
and significances $\Sigma$.}
\medskip
\begin{tabular}{lcccc}
\hline
Decay mode      & Yield &  Efficiency  & ${\cal B}$ & $\Sigma$ \\
                &   & $(\times 10^{-4})$ & $(\times 10^{-4})$ & \\
\hline
$\Bbaro \to D^0 \eta'$  & $49.4\pm 8.7$ & $28.3\pm 0.8$ & 
                                        $1.14 \pm 0.20\,^{+0.10}_{-0.13}$ 
                                        &  $8.9$ \\
$\Bbaro \to D^{*0} \eta'$ & $24.3\pm 6.8$ & $13.0\pm 0.4$ & 
                                        $1.21 \pm 0.34 \pm 0.22$ & $5.3$ \\
\hline
\end{tabular}
\end{center}
\end{table}
%

The major sources contributing to the systematic uncertainty of the
branching fraction measurements are the systematics of the correction
factors for the \de{} and \mb{} shape parameters ($1.9\%$ -- $3.4\%
$, depending on the decay mode considered), 
the choice of the two-dimensional smooth functions ($4.1\%$ -- $12.5\% 
$, this is estimated by fitting the data sample with one-dimensional
functions instead, half the deviation is taken as the error, this
systematic error partly includes the systematics from the PDFs) and
the uncertainty of the reconstruction efficiency of charged tracks ($1.0\% 
$ per track).
Other systematic error contributions are the uncertainty of the 
shape of the PDFs, which
is quantified by varying each parameter of the PDFs by $\pm 1 \sigma$
of its nominal value. The changes in the yield are added in quadrature,
resulting in an error of $0.9\%$ -- $1.8\% 
$. The uncertainties in the
reconstruction efficiencies for all $\eta$, $\pi^0$ and photons
together give a systematic error in the range of $2.9\%$ -- $3.8\%
$. The particle
identification fake rate and efficiency uncertainties for all charged
tracks yield a systematic error of up to $0.4\% 
$. The efficiency is
estimated from MC data samples with an accuracy of $2.5\%$ -- $5.0\% 
$. By
requiring different \LR{} selection criteria the systematic error from
this source is evaluated to be $2.0\%$ -- $8.9\% 
$. The subdecay branching
fraction uncertainties as listed by the PDG~\cite{PDG} result in an
error of $1.5\%$ -- $3.3\% 
$ and the number of \BB{} events that are recorded with the
Belle detector is known to an accuracy of $1\%	
$. The systematic
error contributions described above are added in quadrature and result
in a total systematic error of $^{+8.6}_{-11.5}\%
$ for $\Bbaro \to
D^0 \etap$ and $^{+18.1}_{-18.2}\% 
$ for $\Bbaro \to D^{*0} \etap$.

In summary, we observe the decays $\Bbaro \to D^0 \etap$ and $\Bbaro \to D^{*0}
\etap$ with significances including systematic uncertainty 
of $8.9$ and $5.3$ standard deviations, respectively. 
The latter is observed for 
the first time. Our branching fraction for $\Bbaro \to D^0 \etap$ is about one
standard deviation below another recent measurement~\cite{BaBarD}.
Both branching fractions are higher than early theoretical
predictions but are in agreement with recent work~\cite{Lue03}.

We thank the KEKB group for the excellent operation of the
accelerator, the KEK Cryogenics group for the efficient
operation of the solenoid, and the KEK computer group and
the NII for valuable computing and Super-SINET network
support.  We acknowledge support from MEXT and JSPS (Japan);
ARC and DEST (Australia); NSFC (contract No.~10175071,
China); DST (India); the BK21 program of MOEHRD and the CHEP
SRC program of KOSEF (Korea); KBN (contract No.~2P03B 01324,
Poland); MIST (Russia); MESS (Slovenia); Swiss NSF; NSC and MOE
(Taiwan); and DOE (USA).


\begin{thebibliography}{99}

\bibitem{Belle1}
Belle Collaboration, K.~Abe {\it et al.}, Phys. Rev. Lett.{\bf 
88}, 052002 (2002).

\bibitem{Belle2}
Belle Collaboration, A.~Satpathy {\it et al.}, Phys. Lett. B{\bf 
553}, 159 (2003).

\bibitem{CLEO}
CLEO Collaboration, S.~Ahmed {\it et al.}, Phys. Rev. D{\bf 
66}, 031101 (2002).

\bibitem{BaBarD}
BaBar Collaboration, B.~Aubert {\it et al.}, Phys. Rev. D{\bf 
69}, 032004 (2004).

\bibitem{Neubert}
M.~Neubert and B.~Stech, in {\it Heavy Flavours}, 2nd edition, ed. by
A.J.~Buras and M.~Lindner (World Scientific, Singapore, 1998), p. 294.

\bibitem{Hou01}
C.-K.~Chua and W.-S.~Hou, Phys. Rev. D{\bf 65}, 096007 (2001).

\bibitem{Lue03}
C.-D.~L\"ue, Phys. Rev. D{\bf 68}, 097502 (2003).

\bibitem{Mantry04}
S.~Mantry {\it et al.}, plenary talk, hep-ph/0401058.

\bibitem{CC}
Throughout this paper, 
the inclusion of the charge conjugate mode decay is implied
unless otherwise stated.

\bibitem{KEKB}
S.~Kurokawa and E.~Kikutani, Nucl. Instr. Meth., A {\bf 499}, 1
(2003), and other papers included in this Volume.

\bibitem{QQ}
The event generator for B meson decay ``QQ'' was developed by the CLEO 
collaboration. http://www.lns.cornell.edu/public/CLEO/soft/QQ (unpublished).

\bibitem{GEANT}
R.~Brun, R.~Hagelberg, M.~Hanroul and J.~C.~Lassalle, CERN-DD-78-2-REV (1978).

\bibitem{Belle}
Belle Collaboration, A.~Abashian {\it et al.}, Nucl. Instr. Meth. 
A {\bf 479}, 117 (2002).

\bibitem{PDG}
S. Eidelman {\it et al.}, Phys. Lett. B 592, 1 (2004).

\bibitem{SFW}
 The Fox-Wolfram moments were introduced in
 G.~C.~Fox and S.~Wolfram, Phys. Rev. Lett. {\bf 41}, 1581 (1978).
 The Fisher discriminant used by Belle, based on modified Fox-Wolfram
 moments, is described in 
 K.~Abe {\it et al.} (Belle Collab.), Phys. Rev. Lett. {\bf 87},
 101801 (2001) and
 K.~Abe {\it et al.} (Belle Collab.), Phys. Lett. {\bf B 511}, 151
 (2001). 
     
\bibitem{fisher}
R.~A. Fisher, Annals of Eugenics {\bf7}, 179 (1936).

\bibitem{ARGUS}
The functional form is $x\sqrt{1-x^2}$exp$\left[\alpha(1-x^2)\right]$, where
$x=\mb/\ebeam$. ARGUS Collaboration, H.~Albrecht {\it et al.}, 
Phys. Lett. {\bf B 241}, 278 (1990); {\bf 254}, 288 (1991).

\end{thebibliography}
\end{document}